\documentclass[sigconf,screen,nonacm]{acmart}
\usepackage{graphicx}
\usepackage{xcolor}
\usepackage{soul}
\usepackage{tcolorbox}
\usepackage{caption}
\usepackage{hyperref}
\usepackage{float}
\usepackage{multirow}
\usepackage{url}
\usepackage{multicol}
\usepackage{booktabs}
\usepackage{colortbl}
\usepackage{subcaption}
\usepackage{algorithm}
\usepackage{algpseudocode}
\usepackage[frozencache,cachedir=_minted]{minted}
\definecolor{mintbg}{rgb}{0.95,0.95,0.95}
\setminted{
  bgcolor=mintbg,
  fontsize=\footnotesize,
  breaklines=true,
  frame=none
}
\usepackage{enumitem}
\usepackage[export]{adjustbox}
\usepackage{verbatim}

\newcommand{\toolname}{\textsc{ContextCov}}
\newcommand{\swebenchlite}{\textsc{SWE-bench Lite}}

\usepackage{pifont}

\title{\toolname: Bridging the Gap Between Developer Intent and Autonomous Agent Execution}

\author{Reshabh K Sharma}
\affiliation{
  \institution{University of Washington}
  \city{Seattle}
  \state{Washington}
  \country{USA}}
\email{reshabh@cs.washington.edu}

\begin{document}

\begin{abstract}
As Large Language Model (LLM) agents increasingly execute complex, autonomous software engineering tasks, developers rely on natural language instruction files such as \texttt{AGENTS.md} to express project-specific coding conventions, tooling restrictions, and architectural boundaries. However, because these instructions remain passive text, agents frequently violate documented constraints due to context window saturation or conflicting local context. In autonomous settings without real-time human supervision, such violations rapidly compound into technical debt.

To ground autonomous agents in repository constraints, we introduce \toolname{}, a framework that transforms passive natural language instructions into executable guardrails. Unlike prompt-only or reflection-only compliance approaches, \toolname{} compiles documented constraints into three complementary checks: static AST queries for code patterns, runtime shell shims that intercept prohibited commands, and architectural validators that enforce structural rules. Acting as an automated, continuous reviewer, \toolname{} intercepts agent actions and returns immediate, reproducible violation traces, enabling self-correction before non-compliant changes are finalized.

We evaluate \toolname{} on \swebenchlite{} (12 repositories, 300 tasks). Compared to prompt-only and LLM reflection baselines, \toolname{} achieves 88.3\% constraint compliance (vs.\ 67.0\% and 50.3\%) with 3.4$\times$ lower feedback cost, while maintaining functional correctness. The source code and evaluation results are available at \url{https://github.com/reSHARMA/ContextCov}.
\end{abstract}

\maketitle

\section{Introduction}
\label{sec:introduction}

Large language model (LLM) agents are changing how software is developed. Early systems such as GitHub Copilot~\cite{gh-copilot} primarily acted as autocomplete assistants with near-immediate human verification. Chat-based systems such as Cursor~\cite{cursor} then expanded delegation scope, but developers still supervised execution closely by planning tasks, iterating in chat, and continuously monitoring intermediate reasoning and edits. The current shift is toward autonomous agents. Systems such as GitHub Copilot Workspace~\cite{copilot-workspace}, Devin~\cite{devin}, SWE-Agent~\cite{swe-agent}, and OpenHands~\cite{opendevin} can execute high-level objectives asynchronously, while benchmarks such as SWE-bench~\cite{jimenez2024swebench} accelerate capability development. As autonomy increases, several constraint classes become difficult to verify manually, for example, style violations are buried in large diffs, architectural boundary violations require sustained dependency tracking, and workflow deviations can remain latent until later stages. This shift motivates a fundamental change in how we communicate constraints to agents. The specification can no longer reside solely in the developer's mind or in transient chat history, but must be externalized and persistent.

To close this supervision gap, developers have organically adopted \emph{Agent Instruction files}. Common examples include \texttt{AGENTS.md}, \allowbreak \texttt{CLAUDE.md}, and \texttt{copilot-instructions.md}, which define repository specific rules for autonomous behavior. The \texttt{agents.md} webpage~\cite{agentsmd} reports that more than 60,000 public repositories include such files, though guidance on content quality, update frequency, and enforcement is still evolving. Recent taxonomies identify more than 16 instruction categories~\cite{chatlatanagulchai2025agent}, which we group into three enforcement-oriented classes: \emph{process constraints} govern commands, package managers, and environment setup (e.g., ``Use \texttt{pnpm}, not \texttt{npm}''); \emph{source constraints} govern coding style, naming, and API usage (e.g., ``Prefer arrow functions over anonymous function expressions''); and \emph{architectural constraints} govern module boundaries and dependency direction (e.g., ``The UI layer must not import from the database layer'').

Despite their growing use, instruction files remain passive text rather than executable specifications. Consequently, agents can violate stated constraints when instructions conflict with local context or when long contexts dilute directive salience. These constraint violations appear in both fully autonomous and human-in-the-loop settings because agent edit velocity can exceed human review bandwidth, and many violations especially architectural ones are difficult to detect manually.

Two distinct causes contribute to constraint violations. First, \emph{underspecified instructions} occur when constraints are missing or too ambiguous to operationalize. Second, \emph{agent non-compliance} occurs when constraints are sufficiently specified but generated changes still violate them. This paper focuses on agent non-compliance, because underspecified constraints cannot be enforced reliably without additional author intent. Even when instructions are clear, non-compliance can arise from conflicting repository context, context overload, or capability limits~\cite{liu2026functionalcorrectness, gao2025codehallucination}. Prior work shows that LLMs may over-weight immediate local context relative to higher-level directives~\cite{shi2023large, geng2025controlillusion, hwang2025instructional}, and optimizing for local task completion can produce architectural erosion~\cite{andrews2020erosion} through layer breaches or circular dependencies. These violations can compound over repeated runs and contribute to technical debt accumulation~\cite{molnar2020technicaldebt}.

\toolname{} addresses agent non-compliance by transforming passive instruction files into active, verifiable invariants. We treat instruction files as executable specifications that compile into enforcement checks. \toolname{} operationalizes this through three components: (1)~\emph{Hierarchical Constraint Extraction}, a path-aware parser that preserves scope induced by heading structure. (2)~\emph{Domain-Routed Policy Synthesis}, which routes each extracted constraint to specialized generators for \textsc{Process}, \textsc{Source}, or \textsc{Architectural} enforcement; and (3)~\emph{Multi-Layer Runtime Enforcement}, which combines command interception via PATH injection, static analysis via Tree-sitter queries, and architecture validation via graph analysis and LLM-assisted judgment.

To evaluate this design, we structure our study around four research questions. RQ1 (Extraction Accuracy) asks how accurately \toolname{} extracts enforceable constraints from Agent Instruction files. RQ2 (Enforcement Effectiveness) asks whether \toolname{} reduces constraint violations relative to baselines when all methods use the same fixed five-round feedback budget. RQ3 (Functional Correctness) asks whether enforcing strict constraints negatively impacts the agent's ability to resolve issues. RQ4 (Overhead) asks what the relative cost and latency of executable feedback is compared to LLM reflection.

This paper makes the following contributions:
\begin{enumerate}
    \item We introduce a hierarchical, path-aware algorithm for extracting verifiable constraints from Agent READMEs (Section~\ref{sec:design}).
    \item We present the \toolname{} architecture, comprising domain-specialized code synthesis and a runtime enforcement layer (Section~\ref{sec:design}).
    \item We evaluate \toolname{} on \swebenchlite{} (12 repositories, 300 tasks), measuring extraction quality, constraint compliance, and functional correctness against prompt-only and LLM reflection baselines (Section~\ref{sec:evaluation}).
\end{enumerate}

The remainder of this paper is organized as follows. Section~\ref{sec:example} presents illustrative constraint violation scenarios using Microsoft VS Code's instruction file and shows how \toolname{} detects and mitigates them. Section~\ref{sec:design} details the system design. Section~\ref{sec:evaluation} reports empirical results on \swebenchlite{}. Section~\ref{sec:discussion} discusses implications and limitations. Section~\ref{sec:conclusion} concludes.

\section{Motivating Scenarios}
\label{sec:example}

This section presents three motivating scenarios derived from Visual Studio Code's \texttt{copilot-instructions.md}~\cite{vscode-copilot-instructions} (Figure~\ref{fig:running-example}). We select Microsoft VS Code because it is one of the most widely used and starred open-source software repositories and provides a production-quality Agent Instruction file~\cite{vscode}. We use these scenarios to show how \toolname{} maps natural-language instructions to executable enforcement points across process, source, and architecture levels.

Agents violate constraints for at least two distinct reasons: (1) \emph{instruction omission}, where relevant constraints are not attended to during generation, and (2) \emph{instruction override}, where the agent prioritizes locally common code patterns or short-term task completion over documented project guidance.

The scenarios below are exploratory motivating examples, not controlled benchmark reproductions. To construct them, we manually inspected the instruction file and ran interactive probing conversations with an LLM agent about how it would handle representative tasks and in multiple probes, the agent did not reliably acknowledge or apply relevant instruction-file constraints.
\toolname{} addresses these failure modes by transforming passive instructions into executable checks that provide immediate, actionable feedback to the agent.

\begin{figure}[t]
\begin{minted}[linenos,xleftmargin=2em]{text}
## Validating TypeScript changes
MANDATORY: Always check the `VS Code - Build` watch task
output for compilation errors before running ANY script.

- NEVER run tests if there are compilation errors
- NEVER use `npm run compile` to compile TypeScript files

## Coding Guidelines
### Style
- Use arrow functions `=>` over anonymous function expressions
- Only surround arrow function parameters when necessary.
  For example, `(x) => x + x` is wrong but these are correct:
    x => x + x
    (x, y) => x + y
- Prefer `async/await` over `Promise` and `then` calls

### Core Architecture (`src/` folder)
- `src/vs/base/` - Foundation utilities
- `src/vs/platform/` - Platform services and DI
- `src/vs/editor/` - Text editor implementation
- `src/vs/workbench/` - Main application workbench
  - `workbench/browser/` - Core workbench UI components
  - `workbench/services/` - Service implementations
  - `workbench/contrib/` - Feature contributions
  - `workbench/api/` - Extension host and VS Code API
- **Layered architecture** - from `base` to `workbench`

### Code Quality
- You MUST NOT use storage keys of another component only
  to make changes to that component. You MUST come up with
  proper API to change another component.
\end{minted}
\caption{Excerpt from VS Code's \texttt{copilot-instructions.md}, a production Agent Instruction file used to guide AI coding assistants. Lines 1--6 define process constraints for the TypeScript build workflow, lines 8--15 specify source-level coding conventions, and lines 17--32 establish architectural boundaries and design principles.}
\label{fig:running-example}
\end{figure}

\subsection{Process Violation}
\label{sec:example-process}

Lines 1--6 of Figure~\ref{fig:running-example} specify constraints for the TypeScript build process. Consider an agent tasked with ``Fix the failing editor tests'' that observes test failures in CI logs. Trained on common JavaScript and TypeScript workflows, the agent may run tests directly (for example, \texttt{npm test}) before checking compilation status. However, VS Code uses an incremental watch-based compilation system. Running tests without first verifying compilation can create cascading failures such that the tests fail for reasons unrelated to the target bug, diagnosis becomes noisy, and subsequent patches may address symptoms rather than root causes. If the agent runs \texttt{npm run compile}, which the instructions explicitly forbid, it bypasses the intended workflow and may trigger unnecessary full recompilation. %

\toolname{} addresses this by parsing process constraints and generating checks that intercept shell commands. When an agent attempts to invoke the forbidden \texttt{npm run compile}, it receives immediate, deterministic feedback:

\begin{minted}{text}
$ npm run compile
[ContextCov] BLOCKED: Process constraint violated
  Rule: "NEVER use npm run compile"
  Action: Check VS Code - Build watch task output instead.
\end{minted}

\noindent The agent can then adjust to use the proper workflow instead of proceeding with a forbidden command. More broadly, process violations occur when an agent executes commands that contradict operational constraints. Common patterns include \emph{package manager mismatch} (for example, using \texttt{npm} when \texttt{pnpm} is required), use of non-prescribed build commands, and violation of tool version requirements.

\subsection{Source Violation}
\label{sec:example-source}

Lines 8--15 of Figure~\ref{fig:running-example} specify coding conventions for VS Code. Consider an agent tasked with ``Add a new file watcher callback'' that writes natural-looking TypeScript:

\begin{minted}{typescript}
watcher.onDidChange((event) => {
    this.handleChange(event);
});
\end{minted}

This code is syntactically valid and functionally correct. However, it violates VS Code's style guideline because the single parameter \texttt{event} should not be parenthesized. The agent may have learned this pattern from its pretraining data, where parenthesized single parameters are common in many codebases. Without enforcement, such style inconsistencies can accumulate across the codebase.

\toolname{} addresses this by generating a Tree-sitter query that detects arrow functions with unnecessarily parenthesized single parameters. When the static linter runs, it reports the violation with the source location and a suggested fix, enabling the agent to self-correct before commit:

\begin{minted}{text}
[ContextCov] Source violation: src/vs/workbench/watcher.ts:127
  Rule: "Only surround arrow function parameters when necessary"
  Found: (event) => {...}
  Suggestion: Remove parentheses: event => {...}
\end{minted}

More broadly, source violations occur when an agent generates code patterns that contradict coding standards. These include explicitly forbidden patterns (for example, \texttt{any} types), naming and formatting violations, and deprecated API usage. Agents often introduce these issues by imitating local patterns, even when those patterns conflict with documented standards. %

\subsection{Architectural Violation}
\label{sec:example-arch}

Lines 17--32 of Figure~\ref{fig:running-example} define VS Code's architectural constraints. We distinguish two subtypes: deterministic constraints, which can be verified programmatically, and semantic constraints, which require model-based judgment over intent and responsibility boundaries.

\subsubsection{Deterministic Architectural Constraints}

Lines 17--27 specify the layered architecture and directory structure. Consider an agent tasked with ``Add a utility function for file path normalization'' that must choose where to place new code. The agent might place it in \texttt{src/vs/workbench/common/} because that is where the immediate caller resides. However, VS Code's layered architecture requires such utilities in \texttt{src/vs/base/}, the foundation layer. Similarly, an agent might create a file directly under \texttt{src/vs/workbench/} rather than in allowed subdirectories such as \texttt{browser}, \texttt{services}, \texttt{contrib}, or \texttt{api}. These violations occur when the agent optimizes for local task completion instead of global architectural constraints. %

\toolname{}'s architectural validator constructs a dependency graph and enforces the layered structure. If an agent attempts to create a file in an invalid location, it receives immediate feedback:

\begin{minted}{text}
[ContextCov] Architectural violation: src/vs/workbench/myNewUtility.ts
  Rule: "workbench/ only allows subdirectories: browser,
        services, contrib, api"
  Suggestion: Move to appropriate subdirectory or to base/
\end{minted}

More broadly, deterministic architectural violations include layer violations (dependencies crossing architectural boundaries), cyclic dependencies, and module placement errors. These can be detected using file-path validation and import-graph traversal. %

\subsubsection{Semantic Architectural Constraints}

Lines 29--32 specify a design principle where components must not directly access another component's storage keys but should instead define proper APIs for cross-component changes. Such constraints cannot be verified through file paths or import graphs alone because they require understanding the code's intent. Consider an agent tasked with ``Persist user preferences across sessions'' that discovers another component already stores related settings. Rather than defining a proper API, the agent directly reads and writes to that component's storage keys. This violates encapsulation and couples the components in ways that are difficult to detect in code review.

\toolname{} handles semantic constraints using an LLM-as-judge approach, where a model evaluates whether code changes respect the guideline. These checks produce warnings rather than hard blocks to reduce unsafe false positives in uncertain cases.

\begin{minted}{text}
[ContextCov] WARNING: Possible architectural violation
  Rule: "MUST NOT use storage keys of another component"
  File: src/vs/workbench/contrib/settings/storage.ts
  Concern: Appears to access 'terminal.integrated.fontSize'
           which belongs to the terminal component.
  Suggestion: Define an API in the terminal component instead.
\end{minted}

\section{\toolname{} Design}
\label{sec:design}

\toolname{} consists of two main phases. \emph{Check generation} extracts and synthesizes constraints into executable checks, and \emph{runtime enforcement} executes these checks during agent operation. Given an Agent Instruction file, \toolname{} generates code-based checks that can be incrementally updated as the instruction file evolves. At runtime, source and architectural checks can be enforced over an entire repository, a git diff, or only unstaged changes, while process checks are enforced during agent execution. \toolname{} is implemented in Python. Constraint refinement and intent routing are LLM-assisted steps using OpenAI's \texttt{gpt-5.2-chat}. Final generation for process checks, source checks, deterministic architectural checks, and semantic review rubrics is performed by Claude Code with Opus 4.5 because this setup produces more complete outputs than direct single-shot generation even when file structure context is provided.

\subsection{System Overview}
\label{sec:overview}

\begin{figure*}[t]
\centering
\includegraphics[width=0.6\textwidth,trim=4mm 4mm 4mm 4mm,clip]{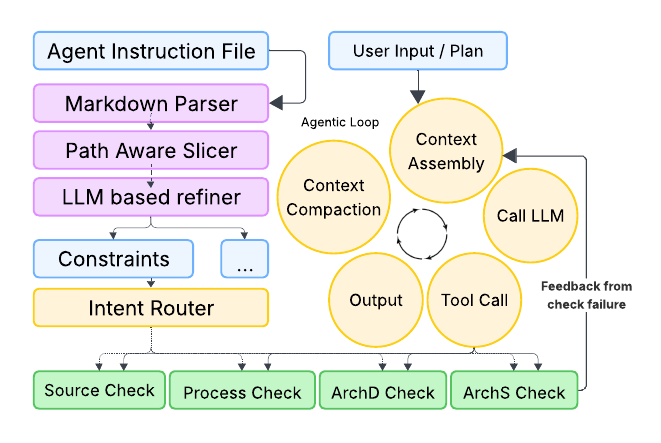}
\caption{Architecture of \toolname{}. The system operates in two phases. First, the Check Generation phase parses Agent Instructions into a Markdown AST, refines constraints via an LLM, and synthesizes executable checks using domain-specialized generators. Second, the Runtime Enforcement phase actively intercepts the agent's shell commands and file modifications, validating actions against the deployed checks and returning deterministic errors as immediate feedback for self-correction.}
\label{fig:architecture}
\end{figure*}

Figure~\ref{fig:architecture} presents the high-level architecture. During extraction, Agent Instruction documents are parsed into a Markdown Abstract Syntax Tree, and a path-aware algorithm traverses the tree extracting constraints as slices with their hierarchical context preserved. An LLM refines each constraint into a standalone, unambiguous statement. During synthesis, each refined constraint is classified by an intent router into one of four enforcement domains and passed to specialized code generators that produce executable Python checks. These checks are stored in a JSON file.

The \textit{Enforcement} phase executes these checks through specialized modules. A Process Interceptor uses PATH shims to intercept shell commands before execution. A Universal Static Linter uses Tree-sitter to analyze source code across multiple languages. An Architectural Validator constructs and queries dependency graphs to detect structural violations. Finally, an LLM-as-judge evaluates semantic architectural constraints that cannot be verified through static analysis alone.

This architecture provides several benefits. When Agent Instructions change, only affected constraints are re-extracted and re-synthesized, enabling efficient incremental updates. Using separate code synthesis modules improves generation quality, as each expert generator is prompted with domain-appropriate patterns and examples. The multiple enforcement modules address the heterogeneity of agent actions. Process constraints require runtime interception, source constraints require AST analysis, deterministic architectural constraints require graph reasoning, and semantic architectural constraints require LLM-based judgment.

\subsection{Constraint Extraction}
\label{sec:extraction}

The extraction phase transforms unstructured Markdown into a set of refined, standalone constraints. This is challenging because Agent Instructions are written for human readers who naturally understand context and scope from document structure. Simple approaches to segmentation fail to preserve \emph{lexical scope}. Consider an excerpt where ``Use pytest'' appears under the headers ``Backend'' and ``Testing''. The instruction's meaning depends on its hierarchical context as it applies to \emph{Backend Testing}, not to the entire project or to Database operations. Extracting this constraint without its context could incorrectly enforce pytest usage for frontend code or database migrations.

\begin{minted}{text}
# Backend
## Testing
- Use pytest
## Database
- Use SQLAlchemy
\end{minted}

To preserve hierarchical context, we parse each Agent Instruction file into a tree structure where internal nodes represent headers at various levels (H1 through H6), leaf nodes represent content blocks (paragraphs, list items, code blocks), and edges represent containment relationships. This tree mirrors how humans mentally organize the document when reading it. For each leaf node representing a potential constraint, we compute its \emph{header path}---the ordered sequence of ancestor headers from the document root to the leaf. In the example above, ``Use pytest'' has the header path \texttt{[Backend, Testing]}. We then use an LLM to rewrite the raw instruction into a standalone, unambiguous statement that incorporates this context, producing a refined constraint like ``For backend testing, use the pytest framework for all unit and integration tests.'' This refinement step disambiguates scope and normalizes diverse phrasings into consistent constraint statements that can be enforced without additional context.

To support incremental updates, we assign each constraint a stable identifier by hashing its header path and content. When a developer modifies one section of an Agent Instruction file, only constraints from that section need to be re-extracted and re-synthesized. Constraints from unmodified sections retain their identifiers and their previously generated checks remain valid, which is important for large instruction files where full re-extraction would be expensive.

\subsection{Constraint to Check}
\label{sec:synthesis}

This phase transforms refined constraints into executable Python code. Rather than using a single monolithic code generator, we route each constraint to a specialized generator based on its enforcement domain. This domain-specific approach improves code quality because each generator is optimized for its particular verification pattern. The Router classifies each constraint into one of four enforcement domains: the \textit{Process Domain} extracts target binary names (e.g., \texttt{npm}, \texttt{yarn}), the enforcement level, and the applicable scope; the \textit{Source Domain} identifies target programming languages, the pattern type to detect, and the severity level; the \textit{Architectural Deterministic Domain} determines the constraint type (dependency direction, cycle detection, or boundary enforcement) and selects an appropriate graph algorithm; and the \textit{Architectural Semantic Domain} preserves the natural language description of the design principle for later LLM evaluation. The router's classification is performed by the LLM, prompted with examples of each domain to improve accuracy.

Each domain has a specialized Expert Generator that produces Python code tailored to that domain's verification needs. The \textit{Process Expert} generates code that inspects command-line arguments and environment variables, returning a verdict that determines whether the command should proceed. The \textit{Source Expert} generates code using Tree-sitter bindings that constructs AST queries and traverses parse trees, returning violations with precise file and line information. The \textit{Architectural Expert} generates code using NetworkX that applies graph algorithms to dependency structures, detecting cycles, layer violations, or boundary crossings. In our implementation, domain-specific outputs are generated through Claude Code with Opus 4.5 rather than by a single direct model call. The generation prompt provides the refined constraint, repository structure, and relevant files, and instructs an iterative self-test loop that samples representative outcomes and revises logic when evidence suggests false positives. Our pipeline also enforces programmatic validation for generated executable checks through syntax checks and dry-run execution before accepting the check, while semantic architectural generation produces reviewer rubrics rather than executable code and uses output-shape validation instead. Tree-sitter grammars are bundled for Python, TypeScript, JavaScript, Go, and Rust, covering the most common languages in repositories with Agent Instructions.

Synthesized checks are stored in a JSON file that can be reviewed and manually edited by developers before deployment, providing a layer of human oversight over the generated checks. The listing below shows an example entry.

\begin{minted}{json}
{
  "constraints": {
    "a7f3b2...": {
      "domain": "PROCESS",
      "original_text": "NEVER use npm run compile",
      "refined_text": "Block npm run compile commands",
      "source_file": "copilot-instructions.md",
      "header_path": ["Validating TypeScript changes"],
      "enforcement_level": "block",
      "check_code": "def check(cmd, args, env): ..."
    }
  }
}
\end{minted}

This separation of synthesis from enforcement provides several advantages. It enables incremental updates where only changed constraints trigger re-synthesis. It allows developers to inspect, modify, or disable specific checks before they take effect. And it ensures fast enforcement at runtime with no LLM calls required, since all checks are pre-compiled Python code.

\subsection{Runtime Check Enforcement}
\label{sec:enforcement}

The enforcement phase executes synthesized checks through four specialized modules, each designed for a particular type of constraint. Enforcing process constraints requires intercepting shell commands before execution, which we accomplish through \texttt{\$PATH} manipulation for simplicity and portability. \toolname{} creates a shim directory containing wrapper scripts for monitored binaries and prepends this directory to \texttt{\$PATH}. Each monitored binary (e.g., \texttt{npm}, \texttt{yarn}, \texttt{pip}) has a corresponding wrapper script:

\begin{minted}{bash}
#!/bin/bash
# .contextcov/bin/npm
exec python -m contextcov.shim npm "$@"
\end{minted}

When a command is executed, the shell resolves the binary via \texttt{\$PATH} and finds our shim first. The shim loads relevant checks from the JSON file and executes each check with the command arguments and environment. If any check fails, the shim prints an error message to stderr explaining which constraint was violated and exits with code 1, preventing the command from running. If all checks pass, the shim uses \texttt{which -a} to find the real binary and executes it with the original arguments, making the interception transparent to the agent.

For source code analysis, \toolname{} uses Tree-sitter~\cite{tree-sitter}, which provides incremental parsing, graceful error recovery, and a uniform API across languages. These properties make it suitable for analyzing code in diverse repositories. The linter scans source files matching configured patterns, parses each file with the appropriate grammar based on file extension, and runs all applicable checks. Each check receives the parse tree, the raw source bytes, and the file path, allowing it to construct queries, traverse the AST, and report violations with precise locations. The following example shows a generated check that detects \texttt{Promise.then()} usage:

\begin{minted}{python}
def check(tree, source_bytes, filepath):
    """Check for Promise.then() usage."""
    query = LANGUAGE.query("""
        (call_expression
          function: (member_expression
            property: (property_identifier) @method)
          (#match? @method "^then$")) """)
    violations = []
    for node, name in query.captures(tree.root_node):
        line = source_bytes[:node.start_byte].count(b'\n') + 1
        violations.append(Violation(filepath, line,
            "Use async/await instead of .then()"))
    return violations
\end{minted}

Deterministic architectural constraints are enforced by constructing and querying a dependency graph. The graph is built by parsing import statements using language-appropriate techniques: Python's \texttt{ast} module extracts \texttt{import} and \texttt{from ... import} statements, Tree-sitter queries extract \texttt{import} declarations and \texttt{require()} calls from JavaScript and TypeScript, and regex-based extraction handles Go import blocks. The resulting graph is stored as a NetworkX \texttt{DiGraph} where nodes represent modules and edges represent dependencies. Architectural checks operate on this graph to detect violations such as layer crossings, cycles, or improper module placement. The following example enforces that files under \texttt{src/vs/workbench/} must reside in one of the allowed subdirectories:

\begin{minted}{python}
def check(graph):
    """Enforce workbench/ subdirectory structure."""
    allowed = {'browser', 'services', 'contrib', 'api'}
    violations = []
    for node in graph.nodes:
        if node.startswith('src/vs/workbench/'):
            parts = node.split('/')
            if len(parts) > 3:
                subdir = parts[3]
                if subdir not in allowed:
                    violations.append(ArchViolation(
                        source=node,
                        message=f"Must be in {allowed}"))
    return violations
\end{minted}

Some architectural constraints cannot be verified through static analysis alone because they require understanding the intent behind code. For example, the constraint ``Do not use another component's storage keys, define a proper API instead'' cannot be checked by examining import graphs or file paths---it requires understanding what constitutes a ``storage key,'' which code belongs to which ``component,'' and whether an access pattern constitutes a violation. For these semantic constraints, we use an LLM-as-judge approach~\cite{crupi2025llmasjudge}. When evaluating code changes, the LLM receives the constraint text, the relevant code snippet, and surrounding context such as file path and neighboring code. The LLM then judges whether the code respects the design principle, providing a verdict and explanation. Semantic checks produce \texttt{WARNING} verdicts rather than hard blocks, flagging potential violations for human review while avoiding false positives from overly strict interpretation.

\subsection{Handling Ambiguity}
\label{sec:failclosed-design}

A key design decision is how to handle ambiguous constraints. Agent Instructions are often written informally, and the same intent can be expressed in many ways. When a constraint like ``Use \texttt{pnpm}'' is encountered, should it block only explicit \texttt{npm} invocations, or also \texttt{yarn} and \texttt{bun}? Should it apply globally or only in certain directories?

We adopt a fail-closed philosophy where ambiguous constraints are interpreted strictly. ``Use \texttt{pnpm}'' triggers blocking of \texttt{npm}, \texttt{yarn}, and \texttt{bun} globally, even though a human reader might interpret this more narrowly. This approach may cause false positives, but we argue that this is the better trade-off for autonomous operation. False positives can be overridden by developers who understand the context, either by refining the Agent Instructions to be more precise or by marking specific checks as exceptions. False negatives allow violations to slip through undetected and accumulate as technical debt. This philosophy aligns with security principles where the system denies the action when uncertainty remains and requires explicit authorization~\cite{saltzer1975protection}.

\section{Evaluation}
\label{sec:evaluation}

We evaluate \toolname{} on the \swebenchlite{} dataset to assess its extraction accuracy and its dynamic impact on autonomous agents. Our evaluation is driven by four research questions:

\begin{itemize}[leftmargin=*]
  \item \textbf{RQ1 (Extraction Accuracy):} How accurately does \toolname{} translate natural-language Agent Instructions into semantically correct executable constraints?
  \item \textbf{RQ2 (Enforcement Effectiveness):} To what extent does active executable feedback reduce constraint violations compared to LLM reflection?
  \item \textbf{RQ3 (Functional Correctness):} Does enforcing strict constraints negatively impact the agent's ability to functionally resolve issues?
  \item \textbf{RQ4 (Overhead):} What is the relative cost and latency of executable feedback compared to LLM reflection?
\end{itemize}

Aggregated statistics are derived from 300 \swebenchlite{} tasks executed across three agent modes.

\subsection{Experimental Setup}
\label{sec:eval-setup}

We evaluate on the 12 repositories comprising the 300 tasks in \swebenchlite{}: \texttt{astropy}, \texttt{django}, \texttt{matplotlib}, \texttt{seaborn}, \texttt{flask}, \texttt{requests}, \texttt{xarray}, \texttt{pylint}, \texttt{pytest}, \texttt{scikit-learn}, \texttt{sphinx}, and \texttt{sympy}. Because historical SWE-bench issues predate modern agentic workflows, we systematically generated repository-grounded \texttt{AGENTS.md} files for all 12 repositories. To ensure construct validity, we used a few-shot LLM generator to extract explicit constraints strictly from existing developer documentation (e.g., \path{CONTRIBUTING.md}), linting configurations (\path{pyproject.toml}, \path{setup.cfg}, \path{.pre-commit-config.yaml}), and CI workflows (\path{.github/workflows/}, \path{pytest.ini}, \path{tox.ini}). We strictly prompted the generator against hallucination, requiring all rules to be supported by existing repository artifacts.

Given these instruction files, \toolname{} generated domain-specific checks (source, process, architectural deterministic, and architectural semantic) and executed them across each repository to identify legacy violations and during the benchmark run to capture live violations.

We use OpenCode~\cite{opencode}, an open-source agentic coding framework, as our agent scaffold. The coding LLM is GPT-5.2 and the LLM Reflection critic uses Claude Opus 4.5. We run all 300 tasks from \swebenchlite{} across three agentic configurations:
\begin{itemize}[leftmargin=*]
  \item \textbf{Vanilla (baseline):} the agent receives \texttt{AGENTS.md} in context but no active execution-time compliance feedback.
  \item \textbf{LLM Reflection (baseline):} a critic LLM reviews candidate patches against \texttt{AGENTS.md} and returns natural-language feedback in a macro-loop.
  \item \textbf{\toolname{}:} the same loop structure, but feedback is produced by executable checks with deterministic violation traces.
\end{itemize}

Feedback-conditioned runs (LLM Reflection and \toolname{}) share a maximum correction budget of five rounds.

\subsection{RQ1: Extraction and Synthesis Quality}
\label{sec:rq1}

Table~\ref{tab:repo-stats} (left) summarizes extraction and synthesis statistics across the 12 repositories. \toolname{} synthesized \textit{788 executable checks} from instruction files across all four domains. Nearly all generated Python checks (780/781) parsed successfully, indicating robust syntactic generation.

To evaluate semantic correctness, we ran checks across repositories, identifying 5,435 potential legacy violations. We uniformly sampled 384 violations (95\% confidence, $\pm$4.72\% margin) for blind human annotation, achieving 82.81\% extraction precision. An LLM-as-judge (Claude Opus 4.5) achieved 85.42\% agreement with human annotators (Cohen's $\kappa = 0.61$), validating its use for subsequent dynamic evaluations.

\begin{table*}[t]
\centering
\caption{Repository-level statistics: extraction \& check synthesis (RQ1, left) and harness resolution (RQ3, right). Van.\ = Vanilla, LR = LLM Reflection, CC = \toolname{}.}
\label{tab:repo-stats}
\small
\setlength{\tabcolsep}{4pt}
\begin{tabular*}{\textwidth}{@{\extracolsep{\fill}} l rrrrrrr | rrrrrr @{}}
\toprule
& \multicolumn{7}{c|}{\textbf{Extraction \& Synthesis (RQ1)}} & \multicolumn{6}{c}{\textbf{Harness Resolution (RQ3)}} \\
\cmidrule(lr){2-8} \cmidrule(l){9-14}
\textbf{Repository} & \textbf{Lines} & \textbf{Seg.} & \textbf{Src} & \textbf{Proc} & \textbf{A-D} & \textbf{A-S} & \textbf{Chk} & \textbf{$n$} & \textbf{Van.} & \textbf{\%} & \textbf{LR} & \textbf{CC} & \textbf{$\Delta$} \\
\midrule
astropy & 151 & 54 & 25 & 17 & 15 & 12 & 69 & 6 & 3 & 50 & 3 & 2 & $-1$ \\
django & 110 & 61 & 35 & 22 & 5 & 11 & 73 & 114 & 68 & 60 & 70 & 77 & $+9$ \\
matplotlib & 157 & 71 & 30 & 24 & 20 & 9 & 83 & 23 & 8 & 35 & 10 & 9 & $+1$ \\
seaborn & 106 & 44 & 15 & 23 & 13 & 0 & 51 & 4 & 2 & 50 & 2 & 2 & 0 \\
flask & 102 & 52 & 25 & 38 & 15 & 1 & 79 & 3 & 1 & 33 & 1 & 1 & 0 \\
requests & 100 & 54 & 16 & 30 & 15 & 6 & 67 & 6 & 6 & 100 & 2 & 6 & 0 \\
xarray & 44 & 17 & 4 & 12 & 1 & 2 & 19 & 5 & 1 & 20 & 2 & 2 & $+1$ \\
pylint & 255 & 130 & 13 & 76 & 27 & 19 & 135 & 6 & 3 & 50 & 2 & 0 & $-3$ \\
pytest & 156 & 55 & 29 & 19 & 13 & 6 & 67 & 17 & 9 & 53 & 9 & 10 & $+1$ \\
scikit-learn & 23 & 12 & 2 & 2 & 0 & 5 & 9 & 23 & 16 & 70 & 13 & 15 & $-1$ \\
sphinx & 104 & 43 & 13 & 17 & 14 & 9 & 53 & 16 & 6 & 38 & 5 & 7 & $+1$ \\
sympy & 190 & 72 & 34 & 16 & 20 & 13 & 83 & 77 & 36 & 47 & 38 & 41 & $+5$ \\
\midrule
\textbf{Total} & \textbf{1,498} & \textbf{665} & \textbf{241} & \textbf{296} & \textbf{158} & \textbf{93} & \textbf{788} & \textbf{300} & \textbf{159} & \textbf{53} & \textbf{157} & \textbf{172} & \textbf{$+13$} \\
\bottomrule
\end{tabular*}
\end{table*}

\subsection{RQ2: Effectiveness of Active Constraint Enforcement}
\label{sec:rq2}

RQ2 asks: \emph{To what extent does active, executable feedback reduce constraint violations and improve correction efficiency compared to LLM reflection?}

We define a \textit{final clean} patch as one containing zero actionable violations upon completion. For \toolname{}, violations are intercepted dynamically in-loop; for baselines, we apply the executable oracle post-hoc. \toolname{} achieves \textbf{88.3\% compliance} compared to 67.0\% (vanilla) and 50.3\% (LLM Reflection). Notably, LLM Reflection performs \emph{worse} than the vanilla agent, suggesting that LLM-based critique introduces more drift than it corrects. A post-hoc taxonomy analysis revealed that all of \toolname{}'s residual violations were limited to source-level checks and the process and architectural violations were fully eliminated by deterministic feedback.

\begin{table*}[t]
\centering
\caption{RQ2: Enforcement efficiency---oracle compliance, feedback rounds, and post-hoc LLM critic analysis (300 instances).}
\label{tab:rq2-enforcement}
\small
\begin{tabular*}{\textwidth}{@{\extracolsep{\fill}} lrrrrrr | lrrrr @{}}
\toprule
& \multicolumn{6}{c|}{\textbf{Oracle Compliance \& Feedback Rounds}} & \multicolumn{5}{c}{\textbf{Post-hoc LLM Critic}} \\
\cmidrule(lr){2-7} \cmidrule(l){8-12}
\textbf{Method} & \textbf{Clean} & \textbf{Rate} & \textbf{Avg.\ viol.} & \textbf{Rounds} & \textbf{Rnds$_\textrm{c}$} & \textbf{Hit cap} & \textbf{Patch source} & \textbf{Clean} & \textbf{$n$} & \textbf{Avg.} & \textbf{Total} \\
\midrule
\toolname{} & 265 & 88.3\% & 0.14 & 1.67 & 1.31 & 29 (10\%) & \toolname{} finals & 30.1\% (88/292) & 292 & 1.15 & 336 \\
Vanilla & 201 & 67.0\% & 0.62 & --- & --- & --- & Vanilla finals & 29.2\% (87/298) & 298 & 1.15 & 344 \\
LLM Reflection & 151 & 50.3\% & 1.01 & 2.85 & 2.30 & 85 (28\%) & LLM Refl.\ finals & 50.3\% (151/300) & 300 & 1.01 & 304 \\
\bottomrule
\end{tabular*}
\end{table*}

\begin{figure}[t]
\centering
\includegraphics[width=\columnwidth]{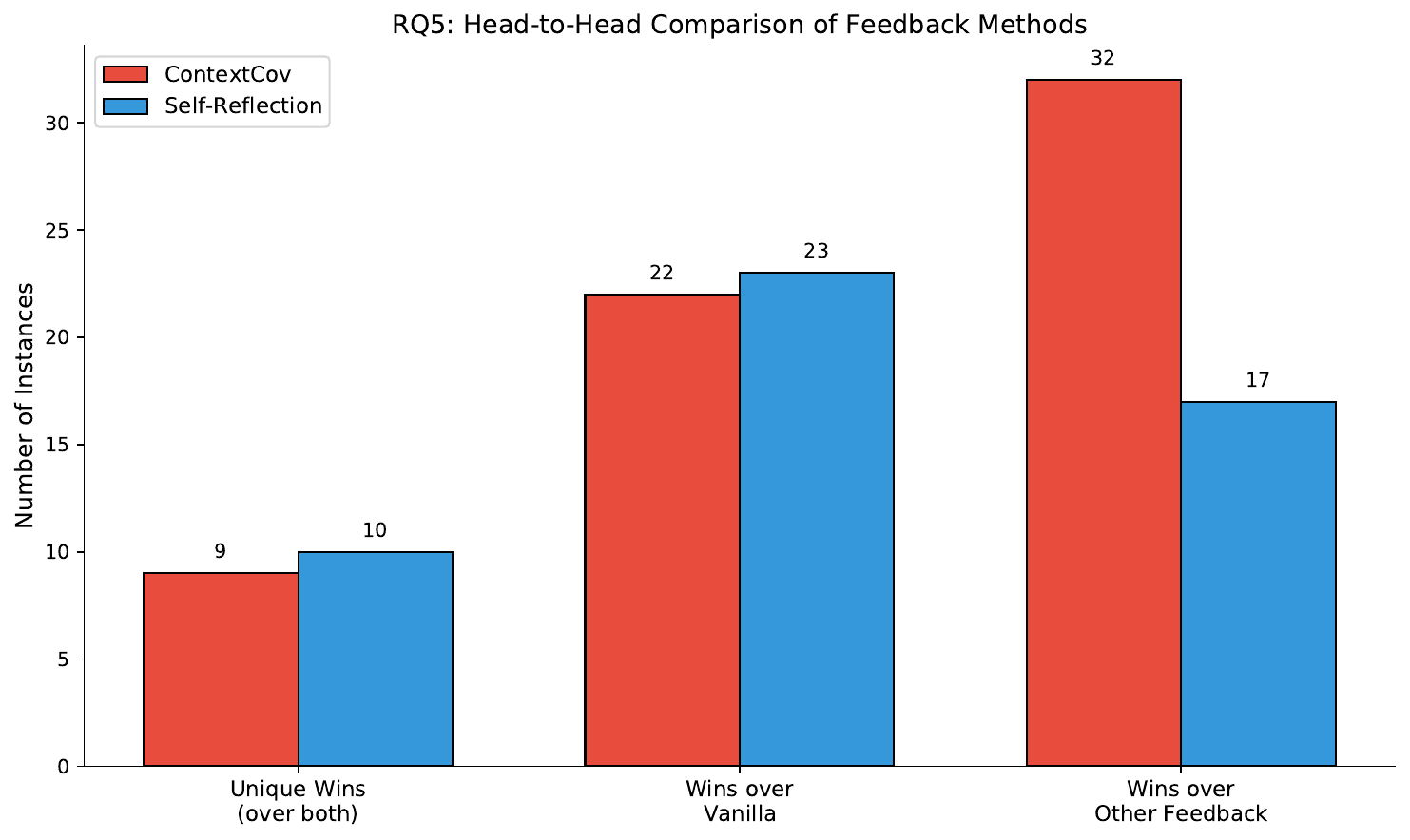}
\caption{RQ2: Head-to-head wins when one method resolves an instance another does not.}
\label{fig:rq2-headtohead}
\end{figure}

\toolname{} also requires fewer feedback rounds to converge (Table~\ref{tab:rq2-enforcement}). LLM Reflection frequently exhausts the five-round cap, exhibiting ``hallucination looping'' where the LLM repeatedly fails to address the same issue. Deterministic error traces allow the agent to pinpoint violations faster. On instances where \toolname{} and LLM Reflection disagreed on final outcomes, \toolname{} won nearly twice as often (Figure~\ref{fig:rq2-headtohead}), confirming that strict enforcement yields better patch trajectories.

\begin{figure}[t]
\centering
\includegraphics[width=\columnwidth]{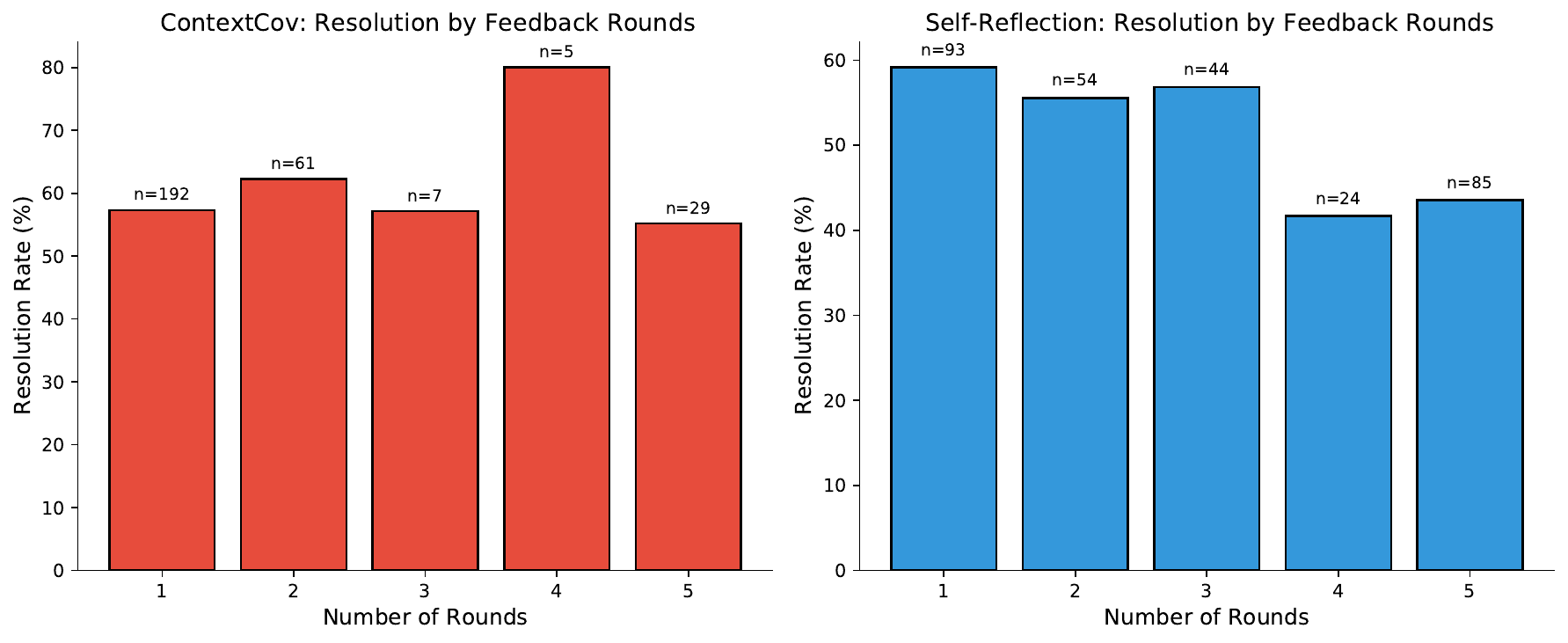}
\caption{RQ2: Harness resolution rate versus number of feedback rounds.}
\label{fig:rq2-rounds}
\end{figure}

We also applied the LLM Reflection critic post-hoc to all final patches (Table~\ref{tab:rq2-enforcement}, right). The critic rates Vanilla and \toolname{} finals similarly, suggesting its complaints are orthogonal to the constraints \toolname{} enforces. Interestingly, LLM Reflection's own finals achieve the highest critic-clean rate, yet the lowest harness resolution (RQ3). This disconnect suggests LLM critics may fixate on stylistic preferences unrelated to functional correctness.

\subsection{RQ3: Impact on Functional Correctness}
\label{sec:rq3}

RQ3 asks: \emph{Does enforcing strict contextual constraints negatively impact the autonomous agent's ability to successfully resolve functional software issues?}

We applied the official \swebenchlite{} evaluation harness to final patches from each method. \toolname{} resolves \textbf{57.3\%} of instances compared to 53.0\% (vanilla) and 52.3\% (LLM Reflection). McNemar's tests on paired outcomes reject the null hypothesis of equal failure rates ($p < 0.05$ for both comparisons), demonstrating that strict constraint enforcement improves rather than hinders functional correctness.

\begin{table}[t]
\centering
\caption{RQ3: Functional correctness---harness resolution, pairwise wins, and McNemar tests ($n{=}300$).}
\label{tab:rq3-functional}
\small
\begin{tabular}{@{}lrrrrr@{}}
\toprule
\textbf{Method} & \textbf{Resolved} & \textbf{Rate} & \textbf{vs Van.} & \textbf{vs other FB} & \textbf{Unique} \\
\midrule
\toolname{} & 172 & 57.3\% & +22 & +32 & 9 \\
LLM Reflection & 157 & 52.3\% & +23 & +17 & 10 \\
Vanilla & 159 & 53.0\% & --- & +25 & 2 \\
\midrule
\multicolumn{6}{@{}l}{\textit{McNemar paired tests (discordant cells: A only / B only)}} \\
\midrule
\multicolumn{4}{@{}l}{Vanilla vs.\ \toolname{}: 9 / 22} & \multicolumn{2}{r}{$p = 0.031$} \\
\multicolumn{4}{@{}l}{Vanilla vs.\ LLM Reflection: 25 / 23} & \multicolumn{2}{r}{$p = 0.885$} \\
\multicolumn{4}{@{}l}{LLM Reflection vs.\ \toolname{}: 17 / 32} & \multicolumn{2}{r}{$p = 0.046$} \\
\bottomrule
\end{tabular}
\end{table}

Table~\ref{tab:rq3-functional} also reports pairwise wins and unique successes. Statistical significance arises because \toolname{}'s wins outnumber its losses, not from isolated outliers. Per-repository analysis (Table~\ref{tab:repo-stats}, right) shows gains concentrated in larger subsets (\texttt{django}, \texttt{sympy}).%

To understand failure modes, we cross-tabulated oracle cleanliness against harness outcomes (Table~\ref{tab:rq3-joint}). For \toolname{}, failures are largely independent of constraint violations and when it fails, the agent simply failed to deduce the bug's logic, not because constraints trapped it. For baselines, oracle-dirty patches correlate with harness failures, suggesting uncorrected constraint violations contribute to functional failure.

\begin{table}[t]
\centering
\caption{RQ3: Oracle cleanliness $\times$ harness resolution ($n{=}300$ per method).}
\label{tab:rq3-joint}
\small
\begin{tabular*}{\columnwidth}{@{\extracolsep{\fill}} llrrr @{}}
\toprule
\textbf{Method} & \textbf{Oracle} & \textbf{Pass} & \textbf{Fail} & \textbf{Total} \\
\midrule
\multirow{2}{*}{\toolname{}} & Clean & 156 & 109 & 265 \\
 & Dirty & 16 & 19 & 35 \\
\midrule
\multirow{2}{*}{LLM Refl.} & Clean & 85 & 66 & 151 \\
 & Dirty & 72 & 77 & 149 \\
\midrule
\multirow{2}{*}{Vanilla} & Clean & 110 & 91 & 201 \\
 & Dirty & 49 & 50 & 99 \\
\bottomrule
\end{tabular*}
\end{table}

Finally, we analyzed the overlap in success across the three methods. Of 300 instances, 127 passed under all three configurations, while 109 failed across the board. Notably, at least one method passed on 191 instances, yielding an implied union success rate of 63.7\%, indicating that while \toolname{} is the strongest individual method, different feedback mechanisms occasionally shepherd the agent down disparate, successful architectural paths.

\subsection{RQ4: Overhead and Cost-Efficiency}
\label{sec:rq4}

\begin{figure}[t]
\centering
\includegraphics[width=\columnwidth]{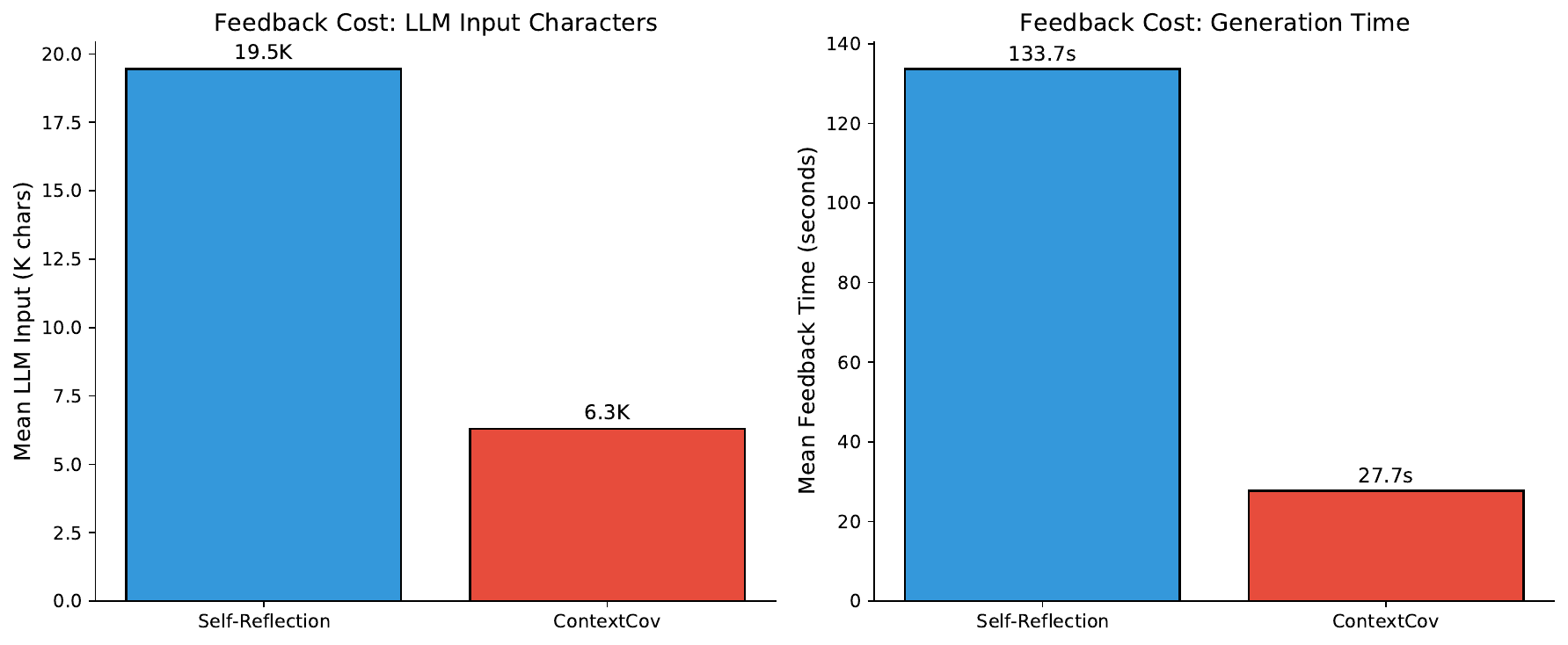}
\caption{RQ4: Mean logged LLM input characters (left) and mean feedback-generation wall time (right) for LLM Reflection versus \toolname{}.}
\label{fig:rq4-cost}
\end{figure}

RQ4 asks: \emph{What are the relative operational costs and latency of deterministic feedback compared to LLM reflection?} We tracked logged characters (as a proxy for token cost) and wall-clock time for feedback generation. As Table~\ref{tab:rq4-cost} and Figure~\ref{fig:rq4-cost} show, \toolname{} reduces feedback cost by $\sim$3$\times$ and latency by $\sim$80\% compared to LLM Reflection. End-to-end time remains dominated by environment setup and test execution.

\begin{table}[t]
\centering
\caption{RQ4: Feedback cost proxies (300 instances).}
\label{tab:rq4-cost}
\small
\begin{tabular*}{\columnwidth}{@{\extracolsep{\fill}} llrr @{}}
\toprule
\textbf{Metric} & \textbf{Method} & \textbf{Mean} & \textbf{Sum} \\
\midrule
\multirow{2}{*}{LLM input chars} & LLM Refl. & 19,455 & 5.84M \\
 & \toolname{} & 6,292 & 1.89M \\
\midrule
\multirow{2}{*}{Feedback time (s)} & LLM Refl. & 133.7 & 40,110 \\
 & \toolname{} & 27.7 & 8,310 \\
\midrule
 & & \textbf{Mean} & \textbf{Median} \\
\cmidrule{3-4}
\multirow{3}{*}{End-to-end (s)} & LLM Refl. & 461.4 & 379.2 \\
 & \toolname{} & 422.4 & 240.0 \\
 & Vanilla & 327.9 & 184.7 \\
\bottomrule
\end{tabular*}
\end{table}

Beyond mean cost, \toolname{} provides more predictable operating costs. LLM Reflection exhibits heavy-tailed dispersion where complex instances cause costs to spiral, while \toolname{}'s variance is substantially lower. Normalizing successful resolutions against feedback cost, \toolname{} achieves 3.4$\times$ higher cost-efficiency than LLM Reflection.

\section{Discussion}
\label{sec:discussion}

The shift from human-supervised to autonomous agents fundamentally changes the role of project documentation. When agents produce changes that violate their instructions, this is not merely a technical bug but a symptom of a deeper mismatch between how we document intent and how agents consume it. We discuss the theoretical and practical implications of treating Agent Instructions as executable specifications.

\subsection{Documentation as Code}

Historically, software documentation was \emph{descriptive}, written for humans who use common sense to bridge the gap between outdated docs and current code. In the era of autonomous agents, documentation like \texttt{AGENTS.md} is \emph{prescriptive}, serving as literal execution context. \toolname{} forces a paradigm shift where documentation must co-evolve with the codebase. By aggressively blocking commands or flagging code that violates Agent Instructions, the tool treats ambiguity and staleness not as minor annoyances but as build-breaking defects. This effectively turns passive Markdown into a living contract between the human developer and the AI agent.

This shift has several implications. When \toolname{} blocks an agent action, developers face a choice. Either the agent's action was wrong and the instructions are correct, or the instructions need updating. Both outcomes improve the repository. This creates an instruction quality feedback loop that does not exist with passive Agent Instructions. Just as code changes undergo testing, Agent Instruction changes should be tested. \toolname{} enables ``instruction tests'' that verify the codebase complies with Agent Instructions before merging instruction changes.

\subsection{The Agentic Feedback Loop}

Because natural language is inherently ambiguous, compiling it into strict checks will yield ``false positives,'' such as globally blocking \texttt{npm} when the developer only meant to block it in \texttt{/backend}. Instead of seeing this as a failure of the tool, we view it as a critical communication mechanism.

When an agent hits a \toolname{} enforcement block, the error is fed back into its context window. This empowers the agent to reason about the failure. If the agent determines the rule was poorly scoped, it can autonomously propose a fix to the Agent Instructions. \toolname{} transforms the agent from a passive consumer of documentation into an active maintainer of project specifications. Our design adopts a strict, fail-closed interpretation of documented constraints (Section~\ref{sec:failclosed-design}). We argue that ambiguity in Agent Instructions is itself a defect. If a rule has valid exceptions, those exceptions should be specified explicitly. By blocking ambiguous cases, \toolname{} forces Agent Instructions to become more precise over time.

\subsection{Executable Interpretability}

A major challenge in prompt engineering is interpretability. A developer writes ``Ensure a clean architecture'' but has no idea how the LLM interprets ``clean.'' \toolname{} addresses this by materializing the agent's understanding into readable, deterministic Python code through Tree-sitter queries and NetworkX graphs. Developers can audit the generated checks to see exactly what the agent thinks the prompt means. If the generated check is too narrow, the developer instantly knows their natural language prompt was too vague. In this sense, \toolname{} acts as a prompt debugger, providing visibility into how natural language constraints are operationalized.

This interpretability enables iterative refinement. When a check produces unexpected results, developers can trace the issue to the original constraint and refine it.

\subsection{Guardrails vs. Sandboxes}

It is important to distinguish \toolname{} from security sandboxes. Security mechanisms like SELinux~\cite{selinux}, AppArmor~\cite{apparmor}, and seccomp~\cite{seccomp} enforce mandatory access controls at the kernel level, while container runtimes like gVisor~\cite{gvisor} and Firecracker~\cite{firecracker} provide process isolation. Runtime verification frameworks~\cite{havelund2001monitoring, chen2007mop, rvmonitor} monitor program execution against formal specifications. \toolname{} is not designed to prevent a malicious actor from compromising a system. Rather, it is designed to prevent a well-intentioned but hallucinating agent~\cite{lin2024hallucination} from degrading the repository. By restricting the execution environment through mechanisms like the Process Interceptor, \toolname{} bounds the agent's action space, ensuring that its exploratory problem-solving stays within approved architectural and operational limits.

\toolname{} assumes three trust boundaries. Agent Instructions are written by authorized maintainers, the repository is not adversarial, and generated checks run without write permissions. Process checks can only block or allow commands, and static checks are read-only AST queries. Organizations should treat \texttt{AGENTS.md} with the same rigor as \texttt{.github/workflows/}.

\section{Related Work}
\label{sec:related}

\toolname{} draws on and extends several lines of prior research. Prior work has studied the ``stale docs'' problem~\cite{aghajani2019software}, API documentation issues~\cite{uddin2015api, robillard2011field, treude2016augmenting}, and README files~\cite{prana2018categorizing, gaughan2025readme}. Techniques detect comment-code inconsistencies~\cite{tan2007icomment, tan2012tcomment, ratol2017detecting} and update comments after code changes~\cite{panthaplackel2020learning, liu2020automating}, while requirements traceability research~\cite{gotel1994analysis, antoniol2002recovering} maintains links between specifications and implementations. However, this prior work focuses on \emph{descriptive} documentation (docstrings, comments) for human developers, whereas \toolname{} targets \emph{prescriptive} documentation (Agent Instructions) where stale docs become execution hazards since agents treat instructions literally.

Research has also translated natural language to formal specifications~\cite{ghosh2016arsenal, brunello2019synthesis, pandita2012inferring, zhong2009inferring}, and specification mining extracts properties from traces~\cite{ammons2002mining, gabel2008symbolic, kremenek2006uncertainty, ernst2007daikon}, while learning-based approaches infer linting rules~\cite{bader2019getafix}. Tools like ESLint~\cite{eslint}, Pylint~\cite{pylint}, Semgrep~\cite{semgrep}, Infer~\cite{calcagno2015infer}, and Error Prone~\cite{aftandilian2012errorprone} enforce standards but require manual configuration or focus on general defects. \toolname{} generates checks from Agent Instructions using domain-routed synthesis, leveraging code search techniques~\cite{digrazia2023codesearch} to produce unified Python scripts operating on Tree-sitter ASTs, NetworkX graphs, and shell shims.

Research on coding agents~\cite{swe-agent, opendevin} has demonstrated autonomous repository navigation and issue resolution, with benchmarks like SWE-bench~\cite{jimenez2024swebench} and BigCodeBench~\cite{zhuo2024bigcodebench} driving rapid progress. Agent alignment frameworks like Constitutional AI~\cite{bai2022constitutional} and Guardrails~\cite{guardrails-ai} focus on general-purpose safety constraints, while research on jailbreaking~\cite{wei2024jailbreak} reveals vulnerabilities in instruction-following. Recent work on detecting agent failures~\cite{sharma2026willful, barke2026agentrx} extracts behavioral rules from prompts and evaluates execution traces post-hoc for compliance violations. \toolname{} complements such analysis by providing \emph{runtime} enforcement: rather than detecting violations after execution, it intercepts non-compliant actions before they are finalized, enabling immediate self-correction.

Finally, tools like Sonargraph~\cite{sonargraph} detect dependency violations, reflexion models~\cite{murphy2001software} compare intended and actual architectures, Knodel and Popescu~\cite{knodel2007architecture} survey architecture compliance checking approaches, and ArchJava~\cite{aldrich2002archjava} embeds architectural constraints directly in the programming language. These tools require architects to manually encode rules in tool-specific formats, whereas \toolname{} extracts architectural constraints from Agent Instructions that developers are already writing, lowering the barrier to architectural enforcement.

\section{Limitations}
\label{sec:limitations}

\toolname{} may flag code that does not actually violate the intended constraint. This can occur when the LLM misinterprets the natural language constraint during extraction, when the synthesized check is overly strict, or when the constraint has implicit exceptions not captured in the Agent Instructions. When false positives occur, they serve as signals for developers to refine their Agent Instruction files, enabling the generation of more appropriate checks over time.

Several validity concerns apply to our evaluation. Generated Agent Instruction files may not perfectly match maintainer intent for repositories that lacked native instruction files, though we mitigate this by grounding generation in repository structure and conventions. Our false-positive analysis relies on a single LLM judge configuration (Claude Opus 4.5), which we partially address through direct human comparison on the same sample and by reporting agreement metrics. \swebenchlite{} includes only 12 Python-heavy repositories and 300 tasks, which may not reflect ecosystems with different languages, build systems, or instruction styles. While the 384-sample design supports 95/5 estimation for proportion metrics, some subgroup analyses may remain underpowered without stratified sampling, and our McNemar tests assume paired instances without correction for multiplicity across the three pairwise comparisons. Broader external validity would additionally require replication on other benchmarks and ecosystems beyond the 12-repository, 300-task Lite split.

\section{Conclusion}
\label{sec:conclusion}

As agents take on more complex, multi-file tasks, certain classes of constraints become impractical to enforce through manual review. Style conventions, architectural boundary rules, and process workflow requirements are inherently difficult to verify by inspection. The specification that once resided in the developer's mind must be externalized to Agent Instruction files, yet these files are passive text, and without automated enforcement, constraint violations accumulate silently. \toolname{} addresses this challenge by treating Agent Instructions as executable specifications, providing automated enforcement for constraints that are difficult to monitor manually. Through hierarchical constraint extraction, domain-specialized code synthesis, and multi-layer runtime enforcement, \toolname{} transforms passive instructions into active invariants that prevent agent violations.

Our empirical evaluation on \swebenchlite{} demonstrates that relying on LLMs to self-police their compliance is both inefficient and error-prone. \toolname{} outperforms both passive prompting and LLM reflection. As autonomous agents become standard contributors to software repositories, frameworks like \toolname{} will be essential for maintaining repository integrity, bridging the gap between what agents are told to do and what they actually do.


\end{document}